\documentclass[11pt,a4paper]{article}
\usepackage{jcappub}
\usepackage{graphicx}
\usepackage{amsmath}

\newlength{\mystretch}
\usepackage{color}
\definecolor{mygrey}{rgb}{0.8,0.8,0.8}
\definecolor{mydarkdarkred}{rgb}{0,0,0}
\definecolor{mydarkdarkblue}{rgb}{0,0,0.7}

\title{Perceiving the equation of state of Dark Energy while living in a Cold Spot}
\author{Wessel Valkenburg} 
\affiliation{Institut f\"ur Theoretische Teilchenphysik und Kosmologie,
RWTH Aachen University, 
D-52056 Aachen, Germany\\
Institut f\"ur Theoretische Physik, Universit\"at Heidelberg,
Philosophenweg 16, D-69120 Heidelberg, Germany} 

\emailAdd{w.valkenburg@thphys.uni-heidelberg.de}

\abstract{The Cold Spot could be an adiabatic perturbation on the surface of last scattering, in which case it is an over-density with comoving radius of the order of 1 Gpc. We assess the effect that living in a similar structure, without knowing it, has on our perception of the equation of state of Dark Energy. We find that structures of dimensions such that they could cause the Cold Spot on the CMB, affect the perceived equation of state of Dark Energy possibly up to ten percent. }

\begin{document}
\maketitle

\section{Introduction}
Since the observation that distant supernovae of type Ia appear dimmer than expected in an Einstein-de Sitter universe,
the cosmological constant has established its return in cosmology~\cite{Riess:1998cb,Perlmutter:1998np}. Attempting to dynamically explain the smallness of the cosmological constant, Dark Energy replaces the cosmological constant with a dynamical theory that drives acceleration of the expansion rate~\cite{Wetterich:1987fm,Sahni:1999gb}. Where the cosmological constant has a constant equation of state, $w=-1$, Dark Energy has an equation of state $w\neq -1$. 

{\color{mydarkdarkred}  Because all our observations of distance measures are necessarily on our past light cone, the accelerated expansion in time is fully degenerate with an accelerated expansion in space: an increasing expansion rate as a function of distance. This is explicated by studies of the large local void scenario (see Refs.~\cite{1995ApJ...453...17M:1995,Tomita:1999qn,Mansouri:2005rf,Alnes:2005rw,Enqvist:2006cg,Enqvist:2007vb,GarciaBellido:2008nz,Clifton:2009kx,February:2009pv,Quartin:2009xr,Marra:2010pg,Moss:2010jx,Biswas:2010xm,Nadathur:2010zm,Marra:2011ct,Bolejko:2011jc,Sussman:2011na} for an incomplete list). In that scenario, Dark Energy is entirely replaced with a large local under-density embedded in a universe that only contains dust and curvature. The price is that the universe is no longer homogeneous, and that we do in fact live in a very special place: the center of a spherically symmetric void that mimics exactly the accelerated expansion caused by a cosmological constant, while locally everywhere the expansion decelerates in time. The size of such a void varies from one to several Gpc, and its center is under-dense with typically $\rho_{\rm center}/ \rho_{\rm outside}\sim 0.1 - 0.3$.

One shortcoming of the void models is that there is as yet no understanding of an origin of such a structure in fundamental physics. If the void is a typical structure in the universe, we would observe many voids outside our own void, notably leaving a strong imprint on the Cosmic Microwave Background (CMB)~\cite{Valkenburg:2009iw}. Nevertheless, these studies prove that inhomogeneities in the matter distribution in the universe can cause the observed distance measures to deviate from the ones in a homogeneous universe.}

{

{\color{mydarkdarkred}
We do not see the kind of spots on the CMB that would indicate the presence of many large voids. What we do see, however, are one or more large Cold Spots in the CMB. One might hence wonder what the physical dimensions of such a perturbation today are, and what influence living inside of such a perturbation could have on our perception of the universe, bearing in mind the degeneracy between accelerated expansion in time and accelerated expansion in distance. The one Spot we have in mind here, has a diameter of roughly 8 degrees and a temperature deviation of rougly $\mathcal{O}(50\sim200)$ $\mu$K \cite{Cruz:2006sv,Zhang:2009qg,Bennett:2010jb}.  It has been considered that the Cold Spot could be caused by an object along the line of sight~\cite{Tomita:2005nu,Inoue:2006rd,Inoue:2006fn,Masina:2008zv}, which we do not address here; we only consider the case of a perturbation on the last scattering surface. 

In Ref.~\cite{Ayaita:2009xm}, Spots with a diameter of 6 degrees and various temperatures are shown to be quite abundant in the CMB, while $\Lambda$CDM predicts an even higher abundance. If there are 100 such Spots on the CMB (which is the case in a $\Lambda$CDM universe if the temperature of the Spots is $\sim 60 \mu$K), then they cover of the order of 8\% of the whole sky. It is hence not unlikely to live in a Spot of this size, with a 6 degrees diameter being slightly smaller than the Cold Spot with an 8 degrees diameter.  
Here we will consider Spots of a comparably small size, a diameter of 5 degrees, although this choice is arbitrary, and a future work could be dedicated to assessing different diameters.

A circle with diameter of 5 degrees on the surface of last scattering, which lies at a comoving angular diameter distance of roughly $d^c_A= 14$ Gpc~\cite{Komatsu:2010fb}, has a comoving radius of roughly 600 Mpc (a diameter of 1.2 Gpc). With that radius, and causing a temperature fluctuation of 50 $\mu$K, the perturbation that is the Cold Spot is today over-dense at the center with roughly $\rho_{\rm center}\simeq 1.1\rho_{\rm outside}$, as we will show later in this work. 

Contrary to the large local void of which the origin is not understood, with a density of $\rho_{\rm center}\simeq 10^{-1} \rho_{\rm outside}$ and a radius of $r\simeq 2$ Gpc embedded in a universe with only dust and curvature, here we consider an existing density perturbation on the surface of last scattering with  $\rho_{\rm center}\simeq  (1 \pm 0.1)\rho_{\rm outside}$ and $r\simeq 600$ Mpc embedded in a $\Lambda$CDM universe. {\em I.e.}, we consider a mixture of accelerating expansion in time and distance. This perturbation in real space may very well be a superposition of gaussian Fourier modes produced during inflation, completely in agreement with the inflationary paradigm~\cite{Ayaita:2009xm}. However, we do not intend to go into that discussion and we refer the reader to for example Refs.~\cite{Ayaita:2009xm,Bennett:2010jb}. All that matters is that the observation of the Spot proves that there is a non-zero probability of living in such spots, of the order of a few percent as mentioned above.
We only wish to argue that studies of the large local void scenario explicate the potential importance of density fluctuations on distance measures, and hence that an undisputably
 realistic perturbation with the same dimensions as the Cold Spot can be expected to have an effect on distance measures in a true $\Lambda$CDM universe. As a simplest approximation, we can model such a rare over-density as a compensated spherical over-density, utilising the LTB metric with a matter distribution such that it matches exactly to the FLRW metric.

}

Smaller voids with sizes of the order of a few hundred Mpc are observed in N-body simulations. Such voids in the presence of a cosmological constant have been considered, assessing the degeneracy between a small local under-density and the value of the cosmological constant~\cite{Marra:2010pg,Amendola:2010ub,Sinclair:2010sb} and the equation of state of Dark Energy~\cite{Romano:2010nc,Romano:2011mx,deLavallaz:2011tj}. In stead, this work focusses on large Cold Spots, which are actually observed in the CMB and which have a much larger physical size but much milder over-density, {\color{mydarkdarkred} and which possibly are in agreement with the inflationary paradigm~\cite{Ayaita:2009xm}, as are the aforementioned smaller voids}.

\begin{figure}
\begin{tabular}{cc}
\includegraphics[width=0.5\textwidth]{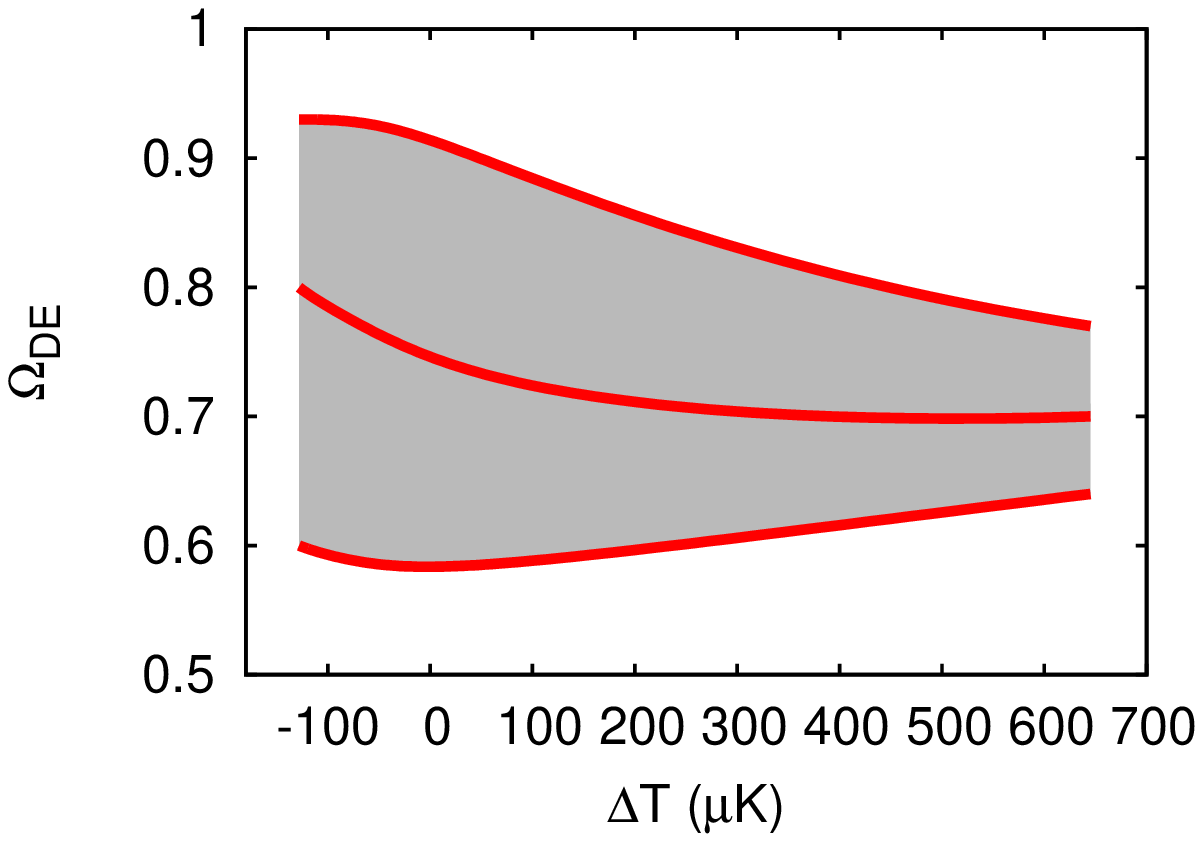}&
\includegraphics[width=0.5\textwidth]{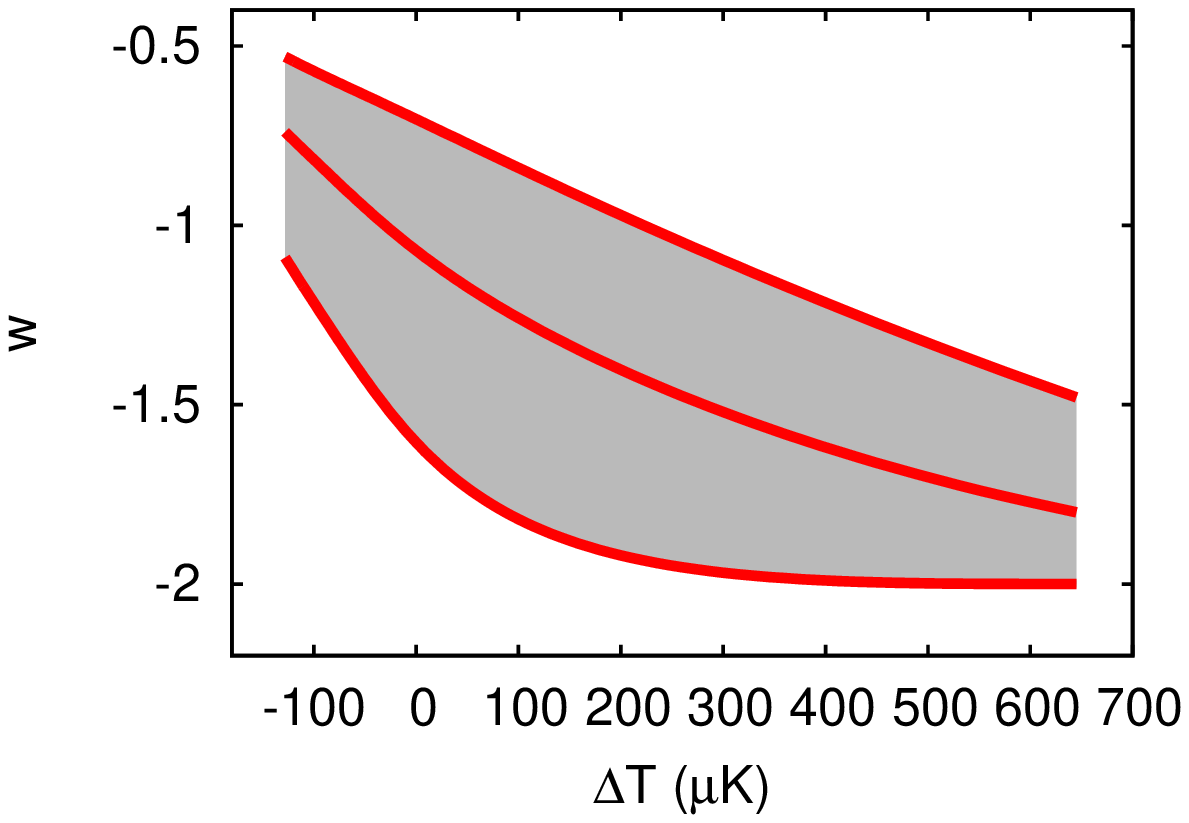}
\end{tabular}
\caption{Central value and 95\% confidence intervals of the marginalized 1D posterior likelihoods of $\Omega_{DE}$ (left) and $w$ (right), as a function of the temperature of the Spot, with an angular diameter of $5^\circ$. That is, the horizontal axis describes the mock data, the vertical axis presents the parameter constraints on a homogeneous cosmology given the mock data.}\label{fig:wlambda}
\end{figure}

We model the Cold Spot using the Lema\^itre-Tolman-Bondi (LTB) metric, assuming that we live in a copy of the object we observe on the surface of last scattering, {\color{mydarkdarkred} embedded in a $\Lambda$CDM universe}. We then construct mock Supernova-data using a numerical module to calculate distance measures in the LTB metric, combining the {\sc VoidDistances} module published in~\cite{Biswas:2010xm} with the exact $\Lambda$LTB~\footnote{With $\Lambda$LTB we denote the Lema\^itre-Tolman-Bondi (LTB) metric with the inclusion of  cosmological constant $\Lambda$. In principle $\Lambda$ has been present in the LTB equations from the beginning, but with $\Lambda$LTB we try to distinguish from theories that explain away $\Lambda$ using the LTB metric with only dust and curvature.} solutions encoded in the module {\sc ColLambda} from Ref.~\cite{Valkenburg:2011tm}. Finally, we fit a homogeneous cosmology with only dust and Dark Energy and no central over- or under-density to the mock SN data, probing the degeneracy between the Cold Spot and $w$, the equation of state of Dark Energy.

Our main result is presented in Figure~\ref{fig:wlambda}, which shows the 95\% confidence level intervals of $\Omega_{DE}$ and $w$ (the equation of state parameter) on the vertical axes, fit against mock data described by the central temperature in a Spot with a diameter of 5$^\circ$ and co-moving radius of 600 Mpc on the horizontal axis. We find that for a Spot with $\Delta T \sim \mathcal{O}(50$ $\mu$K$)$, $w$ is affected at least at the ten percent level, depending on the modeling of the density perturbation, as explained in the body of this paper.

The layout of the work is the following. In Section~\ref{sec:modSpot} we model the Cold Spot in terms of the Lema\^itre-Tolman-Bondi (LTB) metric. In Section~\ref{sec:mockdata} we create mock SN data, given different Cold and Hot Spots. In Section~\ref{sec:res} we fit a homogeneous cosmology with Dark Energy to the mock SN data. We conclude in Section~\ref{sec:conc}.

\section{Modelling the Cold Spot\label{sec:modSpot}}
\subsection{The metric}
We are interested in an order of magnitude estimate of the effect of a over- or under-density with a 600 Mpc radius present in the surface of last scattering, on the temperature anisotropies in CMB photons. We describe the Cold Spot with the Lema\^itre-Tolman-Bondi (LTB) metric,
\begin{align}
ds^2=-dt^2+S^2(r,t)dr^2+R^2(r,t)(d\theta^2+\sin^2\theta\,d\phi^2),
\end{align}
where
\begin{align}
S(r,t)=&\frac{R'(r,t)}{\sqrt{1+2r^2k(r)\tilde M^2}},\\
R(r,t)=&r\,a(r,t),
\end{align}
with the LTB equation,
\begin{align}
\left( \frac{\dot R}{R} \right)^2 =  \left( \frac{\dot a}{a} \right)^2 = & H^2(r,t)=\frac{8 \pi \tilde M^2}{3}\left[   \frac{1}{a^3(r,t)} +  \frac{3k(r)}{4\pi a^2(r,t)}  + \frac{\Lambda}{8\pi\tilde M^2}    \right],\label{eq:LTBi}
\end{align}
valid if the only constituents of the metric are (pressureless) dust, curvature and a cosmological constant. A prime $\,'$ denotes a derivative with respect to radius $r$, an over-dot $\dot \,$ denotes a derivative with respect to time $t$. In Eq.~\eqref{eq:LTBi} we have chosen a gauge for the radial coordinate in which we have
\begin{align}
\int_0^r dr\,M'(r) \sqrt{1+2r^2k(r)\tilde M^2} =&4\pi\tilde M^2 r^3/3,
\end{align}
such that the mass inside a radius $r$ is given by
\begin{align}
M(r)=&4\pi\int_0^r dr\,S(r,t) R^2(r,t) \rho(r,t),\\
M'(r)=&\frac{4\pi\tilde M^2 r^2}{\sqrt{1+2r^2k(r)\tilde M^2}},\\
\rho(r,t)=&\frac{ \tilde M^2 r^2}{R'(r,t)R^2(r,t)}.\label{eq:rhodef}
\end{align}
In this gauge the LTB equation has the form of~\eqref{eq:LTBi}, and the configuration is then described by two functions: the curvature $k(r)$ and the Big-Bang time $t_{BB}(r)$. We choose $t_{BB}(r)\equiv0$, {\color{mydarkdarkred} such that the LTB patch describes a purely growing mode and has no decaying mode}. A shortcoming of this gauge compared to other gauges is that it does not allow for shells of true vacuum with non-zero thickness, where in this gauge $k(r)\rightarrow \infty$. In the presented analysis this is irrelevant however. 

We parametrize the curvature by
\begin{align}
k(r) =& k_{\rm max} W_3\left(\frac{r}{L},0\right) + k_b,
\end{align}
where,
\begin{align}
W_3\left(\frac{r}{L},0\right)=\left\{\begin{array}{lr}
\frac{1}{4\pi^2}\left[ 1 + \pi^2 \left ( 4 - 8 \left(\frac{r }{L}\right)^2 \right) -  \cos \left( 4\pi \frac{r }{L}  \right) \right] & \mbox{ for }  0< r < \frac{L}{2}\\[\mystretch]
\frac{1}{4\pi^2}\left[ - 1 + 8\pi^2 \left ( \frac{r}{L} -1 \right)^2 +  \cos \left( 4\pi \frac{r}{L}  \right) \right]& \mbox{ for }  \frac{L}{2}  < r < L\\[\mystretch]
0& \mbox{ for } r\geq L,
\end{array}\right.\label{eq:theprofile}
\end{align}
is the third order of the function $W_n(x,\alpha)$ which interpolates from $1$ to $0$ in the interval $\alpha < x < 1$, while remaining $C^n$ everywhere, as defined in Ref.~\cite{Valkenburg:2011tm}. Hence $k(r)$ is $C^3$ everywhere, such that the metric is $C^2$ and the Riemann curvature is $C^0$. The constant $k_b$ describes background in which the spherical object is embedded. Since $k_b$ is a constant, the background is an exact Friedmann-Lema\^itre-Robertson-Walker solution, for all $r>L$. {\color{mydarkdarkred} In other words, choosing such an interpolation from a central constant to an outer constant guarantees a mass compensation inside the LTB patch in order to have exact FLRW at $r>L$, without encountering any singular shells.}

Normalizing $a(r>L,t_0)=1$ where $t_0$ denotes today, we have
\begin{align}
k_b\equiv&\frac{4\pi}{3}\frac{\Omega_k(r_*)}{1-\Omega_k(r_*)-\Omega_\Lambda(r_*)}\label{eq:norm1}\\
\tilde M^2 \equiv& \frac{3 H^2(r_*,t_0) - {\Lambda}}{8\pi} \frac{1}{1+\frac{3k_b}{4\pi}}, \\
t_0\equiv& t(a(r_*,t_0)),
\end{align}
where $r_*$ is any $r>L$.
Throughout the rest of this paper we take $\Omega_k=k_b=0$.

We use the solution to Eq.~\eqref{eq:LTBi} described in Ref.~\cite{Valkenburg:2011tm},
\begin{align}
t(a) - t_{BB}(r) =&\frac{1}{\tilde M}\int_0^{a}\frac{\sqrt{\tilde a}\,d\tilde a}{\sqrt{ \frac{8 \pi}{3} +  2 k(r) {\tilde a}  + \frac{\Lambda}{3\tilde M^2}  {\tilde a^3 }}},\label{eq:t_LTB_full}\\
=&\frac{2}{\sqrt{3\Lambda}}\frac{(-1)^{-\frac{9}{2}}}{\sqrt{\prod_{m=1}^3 z_m}} R_J\left(\frac{1}{a}-\frac{1}{z_1},\frac{1}{a}-\frac{1}{z_2},\frac{1}{a}-\frac{1}{z_3}, \frac{1}{a}  \right),\label{eq:t_LTB_full_soln}
\end{align}
where $R_J(x,y,z,p)$ is Carlson's Elliptic Integral of the Third Kind~\cite{1995NuAlg..10...13C}. The parameters $z_i$ are the three (complex) roots of $\frac{8\pi\tilde M^2}{\Lambda}+\frac{6 \tilde M^2 k(r)}{\Lambda} z_i + z_i^3=0$.

\subsection{The surface of last scattering}
Given some $\delta \rho / \rho$, where
\begin{align}
\frac{\delta\rho}{\rho}\equiv\frac{\rho(r=0,t)-\rho(r>L,t)} { \rho(r>L,t)},\label{eq:deltarho_definition}
\end{align}
we invert definition~\eqref{eq:rhodef} using the exact numerical solution from the module published in Ref~\cite{Valkenburg:2011tm}, to obtain a numerical value for $k_{max}$. Then for any $\delta\rho/\rho$ we know $\rho(\vec x,t)$ throughout the history of the universe, given our choice for parameterization of $k(r)$.

Placing the observer at the center of such an object, we numerically solve the radial null-geodesic equations,
\begin{align}
\frac{dt}{dr} &= - S(r,t)\label{eq:geod1}\\
\frac{dz}{dr} &= (1+z)\dot S(r,t),\label{eq:geod2}
\end{align}
to find the time of last scattering, $t_{LSS}\equiv t(z=1089)$. Next, we wish to calculate the temperature fluctuation $\Delta T / T$ induced on the CMB photons by a same object as the one in which the observer resides. We place the object's center on the surface of last scattering. Hence, there are always two identical objects in the universe, one of which surrounds the observer and one of which is centered on the observers surface of last scattering. In the following we will therefore sometimes refer to the temperature and angular size of the Spot in which the observer lives, which are in reality the temperature and angular size of the spot the observer sees on the surface of last scattering, being identical to the structure in which he or she lives.

\subsection{Bardeen potentials}
{\color{mydarkdarkred} Since the objects center is on the surface of last scattering (illustrated in Figure~\ref{fig:intsachswolfe}), and all motion is radial, any motion in the LTB metric on the surface of last scattering is parallel to the surface of last scattering, and orthogonal to the line of sight. This choice simplifies the calculation of the temperature of a Spot, since there is no velocity term.}
To calculate the temperature fluctuation induced by the aforementioned object, it is easiest to perform a gauge transformation from the synchronous gauge to a gauge closer to the newtonian gauge, in order to calculate the Bardeen potentials. We follow the (linear) methods in Refs.~\cite{Biswas:2007gi,VanAcoleyen:2008cy}. For completeness we describe explicitly the steps from Section 4.1 in Ref.~\cite{Biswas:2007gi}. In general a metric with only scalar perturbations can in spherical coordinates be written as
\begin{align}
ds^2= a^2(\tau) \left\{ -(1+2\psi) d\tau^2 + 2\omega'(r) dr d\tau + \left(1-2\phi + \frac{2}{3}\mathcal{E} \right)dr^2  + \left(1-2\phi - \frac{1}{3} \mathcal{E} \right) r^2d\Omega^2\right\},
\end{align}
with $\mathcal{E}\equiv \chi'' - \frac{\chi'}{r}$.

Comparing to the LTB metric,
\begin{align}
ds^2=-dt^2+S^2(r,t)dr^2+R^2(r,t)(d\theta^2+\sin^2\theta\,d\phi^2),
\end{align}
we see that 
\begin{align}
\psi=&0,\\
\omega'(r)=&0,\\
\mathcal{E}=&\frac{S^2(r,t)}{a^2(r>L,t)}-\frac{a^2(r,t)}{a^2(r>L,t)},\\
\phi=&\frac{1}{2} - \frac{1}{3} \frac{a^2(r,t)}{a^2(r>L,t)} - \frac{1}{6}\frac{S^2(r,t)}{a^2(r>L,t)},\\
d\tau^2=&\frac{dt^2}{a^2(r>L,t)},
\end{align}
where it is understood that $a(r>L,t)$ is a pure function of $t$.

Using the exact semi-analytical solutions to the LTB metric, we can solve the second order differential equations for $\mathcal{E}\equiv \chi'' - \frac{\chi'}{r}$, $ \mathcal{\dot E}\equiv \dot \chi'' - \frac{\dot \chi'}{r}$ and $ \mathcal{\ddot E}\equiv \ddot \chi'' - \frac{\ddot \chi'}{r}$. It can easily be shown that the boundary condition for the solution $\chi(r,t)$ is given by the fact that the only possible solution at $r>L$ is $\chi=\chi'=\chi''=0$.  Then we have all the ingredients necessary to calculate the Bardeen potentials for our scenario (ignoring terms that are zero),
\begin{align}
\Psi=&\frac{1}{2}\left(\partial^2_{\tau} \chi + \partial_\tau \chi \frac{\partial_\tau a(r>L,t)}{a(r>L,t)}\right),\\
\Phi=&\phi + \frac{1}{6}\left( \chi'' + \frac{2\chi'}{r}  \right) + \frac{1}{2}\partial_\tau\chi \frac{\partial_\tau a(r>L,t)}{a(r>L,t)},
\end{align}
where $\partial_\tau \equiv a(r>L,t)\partial_t$. The reason we use both $\Phi$ and $\Psi$, is that they must be equal for our scenario, and therefore they are a useful check of the calculations. Another check we performed is that $\Phi$ and $\Psi$ are constant in time, which they should be for linear matter perturbations in a matter dominated universe.

\subsection{The resulting Spot}
We assume that the 1.2 Gpc sized object (600 Mpc radius) is an adiabatic perturbation{\color{mydarkdarkred}, which is possible since there is no decaying mode ($t_{BB}=0$ everywhere), and we assume that the surface of last scattering goes through the center of the spherical object}. In that case, the temperature fluctuation is given by the ordinary Sachs-Wolfe effect~\cite{Sachs:1967er,Durrer:2001gq},
\begin{align}
\frac{\Delta T}{T} =& \frac{\Phi}{3}.
\end{align}

The observed Cold Spot has roughly a diameter of 5 degrees $\simeq 0.087$ radians.  The temperature deviation is rougly $\mathcal{O}(50\sim200)$ $\mu$K~\cite{Cruz:2006sv,Zhang:2009qg,Bennett:2010jb}. The comoving angular diameter distance to the last scattering surface in $\Lambda$CDM is $d^c_A = 14$ Gpc per radian~\cite{Komatsu:2010fb}, where $d^c_A(z)=(1+z)d_A(z)$ with $d_A(z)$ the true angular diameter distance. So the diameter $d$ of the spherical object centered on the surface of last scattering, that gives rise to the Cold Spot in the CMB is roughly 
\begin{align}
d \simeq 14 \mbox{ Gpc} \times 0.087 \simeq 1.2 \mbox{ Gpc}.
\end{align}

\begin{table}
\begin{center}
 \begin{tabular*}{0.3\textwidth}{@{\extracolsep{\fill}}l|r}
Parameter & Value\\
\hline
$\Omega_b h^2$ & 0.0253\\
$\Omega_{dm} h^2 $ & 0.1122\\
$H_{0, {\rm outside}}$ & 74.2\\
$w$ & -1
\end{tabular*}
\end{center}
\caption{The parameters used to describe the background cosmology, on top of which the 1.2 Gpc sized structure is imposed. That is, these parameters describe the cosmology for $r>L$. As usual, $h=H_0/ (100$ km / s / Mpc). We have $\Omega_\Lambda = 1 - \Omega_m = 0.75$. }\label{tab:pars}  
\end{table}

For simplicity we hence take the co-moving radius of our object to be 600 Mpc, where the comoving radius is defined as $L_c=\int_0^L \left[S(r,t_0) / S(r_{\rm obs},t_0) \right] dr$ with the observer at $r_{\rm obs}=0$. As a background cosmology, that is for $r>L$, we take the parameters listed in Table~\ref{tab:pars}. Then for different density contrasts today for the structure in which the observer lives, we can calculate the temperature fluctuation and the density contrast that an identical structure has at a distance at $z=1089$ from the observer, taking into account that the structure in which the observer lives slightly affects the angular diameter distance to $z=1089$. 
 Hence, there are two identical objects in the considered model universe: one centered around the observer and one centered on the surface of last scattering. 
The results for some example values of $\delta \rho / \rho (t_0)$ are given in Table~\ref{tab:temps}. Since the observer lives in the center of an identical Spot, the observer's ruler depends on the local scale factor in the Spot. In an under-dense region, which expands faster than its FLRW surrouding, the observer therefore has a larger ruler. This means that for that observer a 600 Mpc structure corresponds to a larger physical structure on the surface of last scattering than for an observer in an over-dense region. Similarly, the surface of last scattering is at a smaller distance for the under-dense observer. This explains why in Table~\ref{tab:temps} the angular sizes of the Spots vary, while their comoving radius is fixed to 600 Mpc.

We define the angular radius of the Spot, $L_{\rm deg}$ as 
\begin{align}
L_{\rm deg} =& \frac{600 \mbox{ Mpc}}{d_A^c(z=1089)} \frac{\mbox{degrees}}{\mbox{radian}}\nonumber\\
\simeq & 31.3^{\circ}  \frac{S(r_{\rm obs},t_0)}{ R(z=1089) / (1 \mbox{ Mpc})},
\end{align}
with the co-moving angular diameter distance 
\begin{align}
d_A^c(z) = (1+z) d_A(z) = (1+z) R(r(z),t(z)) / S(r_{\rm obs},t_0),
\end{align}
 and $r_{\rm obs}=0$.

\begin{table}
\begin{tabular*}{\textwidth}{@{\extracolsep{\fill}}rrrr}
$\frac{\delta\rho}{\rho}(t_0) $ & $\frac{\delta\rho}{\rho}(z=1089) $  & $ \Delta T $ ($\mu$K) & Spot diameter \\  \hline
 $      5\times 10^{-1}$ & $      4.41\times 10^{-4}$ &        -127.96 &           4.32$^{\circ}$ \\
 $      3\times 10^{-1}$ & $      2.98\times 10^{-4}$ &         -95.08 &           4.53$^{\circ}$ \\
 $      1\times 10^{-1}$ & $      1.14\times 10^{-4}$ &         -40.63 &           4.79$^{\circ}$ \\
 $      1\times 10^{-2}$ & $      1.22\times 10^{-5}$ &          -4.61 &           4.93$^{\circ}$ \\
 $     -1\times 10^{-2}$ & $     -1.24\times 10^{-5}$ &           4.75 &           4.96$^{\circ}$ \\
 $     -1\times 10^{-1}$ & $     -1.34\times 10^{-4}$ &          54.67 &           5.13$^{\circ}$ \\
 $     -3\times 10^{-1}$ & $     -4.91\times 10^{-4}$ &         237.19 &           5.57$^{\circ}$ \\
 $     -5\times 10^{-1}$ & $     -1.07\times 10^{-3}$ &         645.17 &           6.24$^{\circ}$ \\
\end{tabular*}
\caption{Density contrast as in Eq.~\eqref{eq:deltarho_definition} at the last scattering surface (second column) and induced temperature perturbation relative to the CMB temperature (third column) for different density contrasts today (first column). The cosmology is described by the parameters given in Table~\ref{tab:pars} plus an LTB patch with a radius of 600 Mpc (diameter of 1.2 Gpc), with the given density contrast today and a central observer. Hence, the universe contains two identical LTB patches, one centered on the observer, one on the surface of last scattering. The angular size of the Spot, defined as size $\equiv 2 L_{\rm deg}$, varies as a function of density contrast, because the size of the observer's ruler changes with the density. An observer in an underdensity has a larger ruler, such that the 600 Mpc patch corresponds to a larger fraction of the distance to the last scattering surface. The temperature of the CMB today is 2.726 K~\cite{Bennett:1996ce}.}\label{tab:temps}
\end{table}

\begin{figure}
\begin{tabular}{cc}
\includegraphics[width=0.48\textwidth]{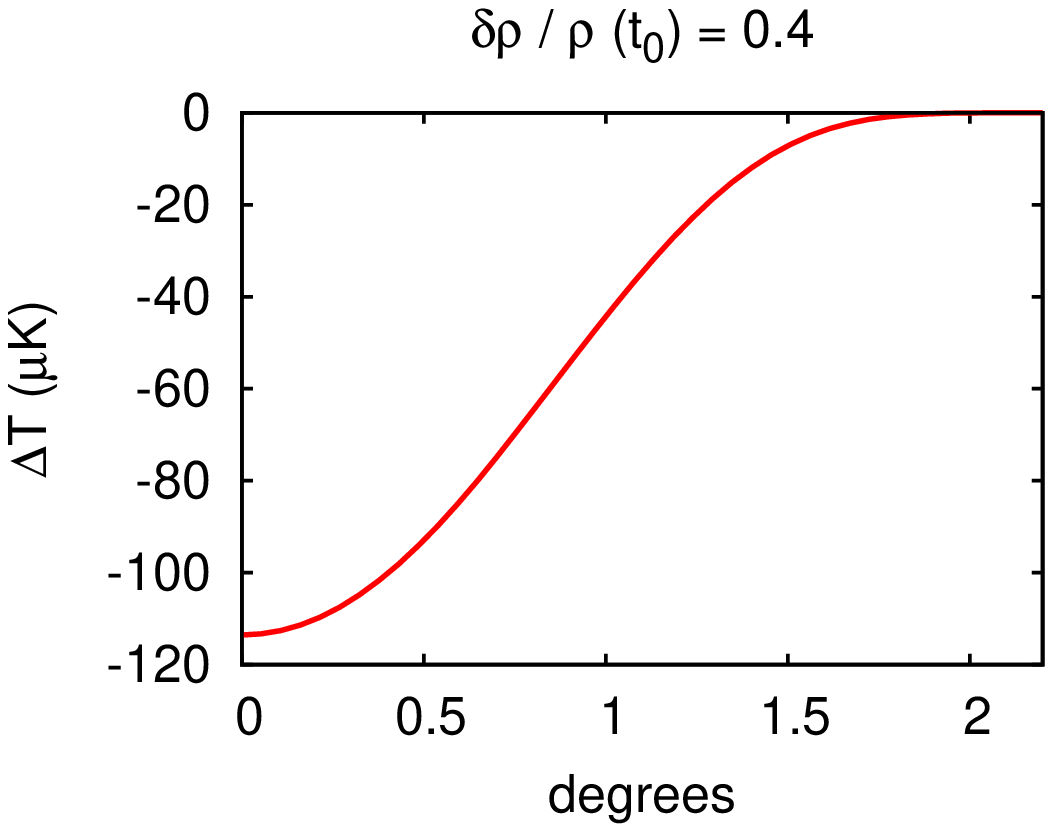}&
\includegraphics[width=0.52\textwidth]{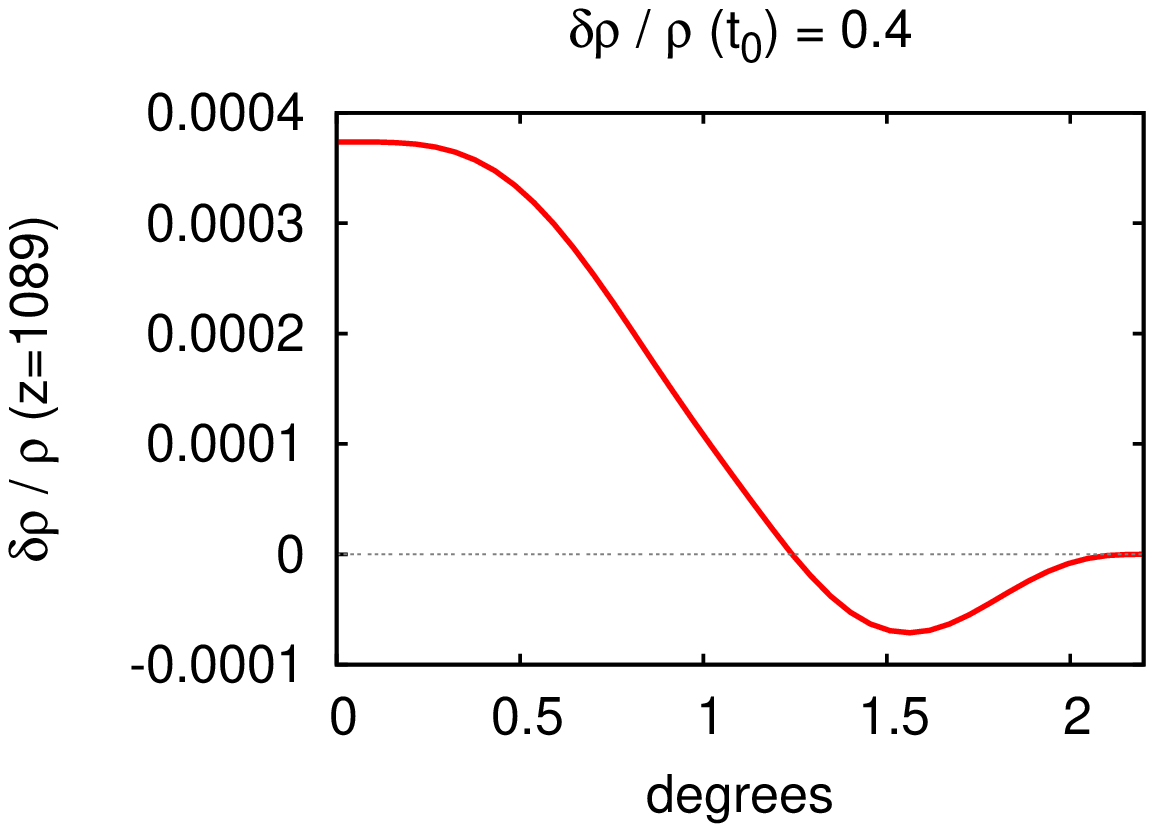}\end{tabular}
\caption{{\em Left:} Temperature fluctuation as a function of angular distance from the center of the object, in this case a 5$^\circ$ circular coldSpot which today would be a spherical over-density of $\delta \rho / \rho = 0.4$ with a radius of 600 Mpc. {\em Right:} The over-density profile of the same over-density as in the left figure, at the time of decoupling.}\label{fig:tempprofile}
\end{figure}

In Figure~\ref{fig:tempprofile} we show the temperature as a function of the angular distance from the center. The temperature interpolates smoothly from the central minimum up to zero on the edge. Even though the edge has an under-density due to the compensated way of modeling the over-density, the gravitational potential in the under-dense shell is, relative to the homogeneous surrounding medium, still lower than the surrounding zero potential. Hence, the under-dense ring does not produce a hot ring around the Cold Spot. The reason is that we only `see' the potential. The under-dense shell is closer to the center of the potential compared to its surroundings. Hence not surprisingly, it resides at a lower value of the potential. Whether this absence of  a ring is physical or not, should be subject to further investigation. Anyway, for exactly compensated over-densities (with an always negative gravitational potential), the ordinary Sachs-Wolfe effect predicts {\em no} hot ring, although a non-linear integrated Sachs-Wolfe effect may. The converse also holds, the ordinary Sachs-Wolfe effect predicts no cold ring around a Hot Spot. If such rings are observed, that could be a hint that the origin of such Spots is the late time integrated Sachs-Wolfe effect, or that they are non-linear perturbations~\cite{Granett:2008ju,Valkenburg:2009iw,Inoue:2010rp}.

{\color{mydarkdarkred}In the large local void scenario, the observer is confined to live close to the center of the void, because the radial velocity of matter scales with the distance from the center, and only close to the center the dipole is in agreement with the dipole of 3.355 mK as observed by WMAP. In the case considered here, with the observer living in an object which can cause the Cold Spot such as the example case of Figures~\ref{fig:tempprofile}~and~\ref{fig:intsachswolfe}, the maximum dipole caused by radial motion due to living off-center is 12 mK, a peak which lies around $r\simeq L/2$. Such a structure hence is hardly constrained by the WMAP dipole observation.}

\subsection{Shortcomings\label{sec:shorts}}
There are two important assumptions, that can be considered shortcomings in our modeling of Spots.  \begin{itemize}
\item We make the crude approximation that the perturbation is embedded in a fully matter dominated universe, all the way back to the photon decoupling. Taking into account the presence of photons, the gravitational potential should still decay slightly at photon decoupling. Ignoring this decay leads to an overestimation of the density perturbation.
\item We assume that the intersection of the last scattering surface and the spherical object goes through the center of the object. That is, the line of sight to the center of the object, hits the last scattering surface at $r=0$ in the object. In Figure~\ref{fig:intsachswolfe} we illustrate the possibility that the object intersects the last scattering surface at another radius. In that case, the matter in the object on the surface of last scattering, will have a non-zero effective velocity with respect to the CMB rest frame, parallel to the line of sight, as explained in the caption of Figure~\ref{fig:intsachswolfe}. In the case of an over-density, which gives a Cold Spot, the matter is falling in, such that it gets an additional Doppler redshift or blueshift, which is ignored since we only account for the ordinary Sachs-Wolfe effect by considering the Bardeen potentials in stead of the explicit geodesic integration. In Figure~\ref{fig:intsachswolfe} one can see that the Doppler term can dominate over the ordinary Sachs-Wolfe effect, and hence that what is a Cold Spot in this paper can even be observed as a Hot Spot when it intersects the surface of last scattering at a different radius. We leave such complications of the model to future work though. Note that the velocity is a monotonic function of radius since the curvature is a monotonic function of radius, and hence that an intersection at different radii will not induce hot or cold rings through the Doppler shift.
\end{itemize}
Both effects should be more carefully considered in order to give more definite estimates of the relation between the temperature of a Spot and the density inside it.

\begin{figure}
\begin{tabular}{cc}
\includegraphics[width=0.45\textwidth]{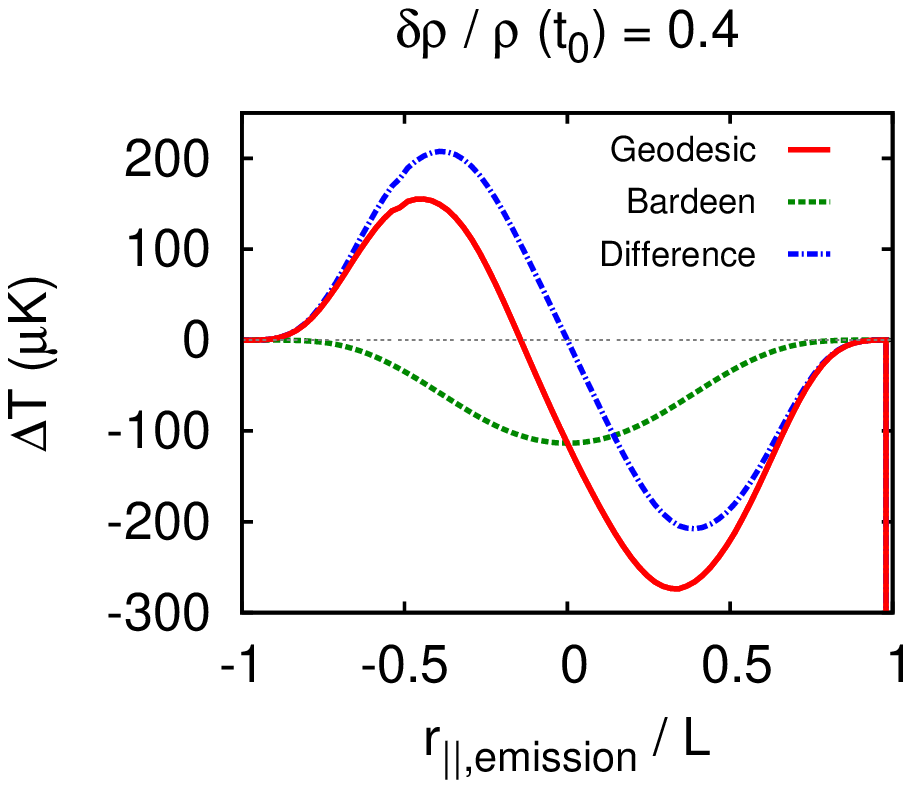}&
\includegraphics[width=0.45\textwidth]{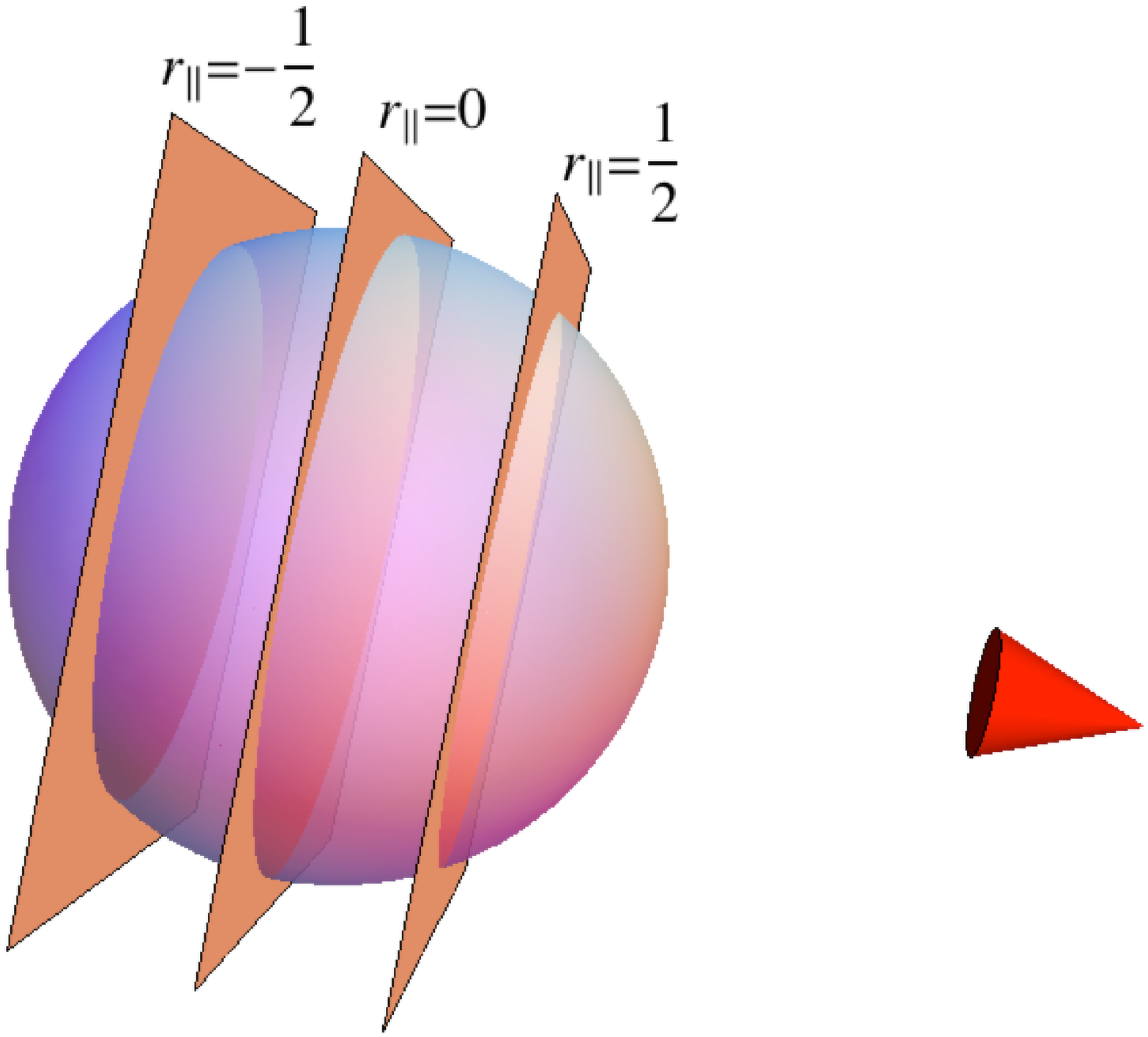}
\end{tabular}
\caption{{\em Left:} The central temperature of the Cold Spot as a function of the radius  $r_{||, {\rm emission}}$, chosen parallel to the line of sight, at which the surface of last scattering intersects the sphere. In solid red we show the temperature fluctuation as calculated by explicit solving of the geodesic equations, which takes into account both the Bardeen potential (ordinary Sachs Wolfe effect, dashed green curve) and the velocity along the line of sight (the dashed-dotted blue curve, which is obtained by taking the difference between the solid red and dashed green curves). {\color{mydarkdarkred} The velocity is automatically taken into account since the coordinates are comoving with the matter, and we assume emission from a fixed coordinate. }
The Bardeen-potential curve is identical to the curve shown in Fig~\ref{fig:tempprofile}, the difference being that in that figure the Bardeen potential is the only contribution to the temperature fluctuation, since the surface of last scattering there intersects the sphere at $r_{||, {\rm emission}}=0$. The fact that in this figure the difference curve (blue dashed dotted) is anti-symmetric in $r_{||, {\rm emission}}=0$, provides a good cross-check between the different calculations of the temperature of the Spot.
{\em Right:} A schematic presentation of three different radii at which the last scattering surface (the three upright planes) intersect the spherical object. The cone on the right of the illustration represents the observer. }\label{fig:intsachswolfe}
\end{figure}

\section{Mock Supernovae\label{sec:mockdata}}
In spherical coordinates and for a central observer, the angular diameter distance $d_A\equiv ds / d\theta$ with $s$ the physical diameter of the object and $\theta$ the angle under which the object is oberved, is given by the metric, $d_A(r,t)=R(r,t)$. To obtain the angular diameter distance to a point on the past light cone, $d_A(z)$ with $z$ the redshift\footnote{In non-spherical cosmologies there is a direction dependence, such that one could write $d_A(\vec n, z)$.}, one needs the pair $\{r,t\}$ corresponding to a certain redshift. Note that the fact that we need the pair $\{r,t\}$, and not only $t$, is the reason why large local voids are considered to replace $\Lambda$ in the first place. Hence, one solves the geodesic equations~(\ref{eq:geod1},\ref{eq:geod2}) backwards in time, starting from the observer back to some redshift $z$. With the observer at $z=0$, the co-moving angular diameter distance becomes $d_A^c(z)\equiv(z+1)d_A(z)$, relating the angular size of an object to its size in the orthonormal coordinates of the observer, if it had expanded along with a fictitious homogeneous background expansion characterized by the redshift $z$.

We show the resulting distance moduli, $\mu \equiv 5 \log_{10}\left[ (1+z)^2 H_0 d_{\rm A}(z) \right] + 25$, for different Spots in Figure~\ref{fig:mocksn}. We generate mock data in the same way as done for example in Ref.~\cite{Park:2010xw}, by taking existing supernovae with their error bars, and replacing their distance moduli with the theoretical distance moduli for a given cosmology with the observer in a given Spot. For completeness, we generate sets using the SDSS Supernovae~\cite{Kessler:2009ys} processed with both the MLSCS2k2 and SALT-II lightcurve fitters, and the Union2 data set~\cite{Amanullah:2010vv} which is processed with the SALT-II lightcurve fitter as well. 

\begin{figure}
\includegraphics[width=0.8\textwidth]{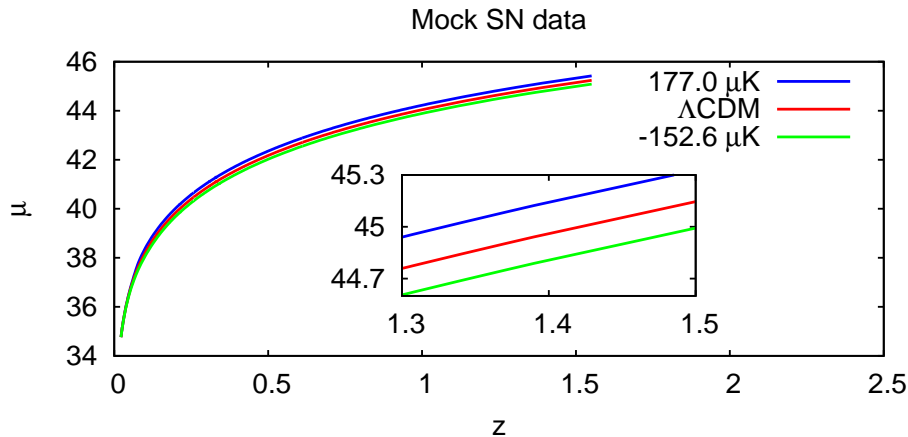}
\caption{Distance modulus $\mu$ versus redshift $z$ for three different mock datasets: for an observer in a Hot Spot of 177.002 $\mu$K (blue line, top), an observer in true $\Lambda$CDM (red line, middle) and an observer in a Cold Spot of -152.610 $\mu$K (green line, bottom). All moduli are normalized to an arbitrary expansion rate, and such that the three lines coincide for the lowest redshift supernova (the far left of the figure). The inset shows the same lines, zooming in on the high redshift range, to make more visible the effect of the different structures on the high redshift distance moduli. Error bars of individual supernovae are not shown, for clarity.}\label{fig:mocksn} 
\end{figure}

\section{Results\label{sec:res}}
Using the mock data described in the previous section, we fit a $w$CDM cosmology with an equation of state for Dark Energy allowed to differ from $w=-1$. The Dark Energy is then parameterized by 
\begin{align}
\rho_{DE}(t) =& H_0^2 \Omega_{DE} \left( \frac{a_0}{a(t)}\right)^{3(1+w)}.
\end{align} 
To perform the fits, we use {\sc cosmomc}~\cite{Lewis:2002ah} for the Monte-Carlo sampling and {\sc camb}~\cite{Lewis:1999bs} for the distance calculations in the FLRW cosmology. 
We allow the following parameters to vary, with their prior limits:
\begin{align}
0 < \Omega_m h^2 < 1,\nonumber\\
20 < H_0 < 100,\nonumber\\
-2 < w < 0,
\end{align}
where $\Omega_{DE}=1-\Omega_m$, and $h=H_0/100$ km s$^{-1}$ Mpc$^{-1}$.

In Figure~\ref{fig:1dpostcurv} we show the marginalized 1-dimensional posterior probability distributions of $\Omega_{DE}$ and $w$ when fitting mock datasets based on the real datasets as described in Section~\ref{sec:mockdata}, while simulating universes with an observer living in the Spots listed in Table~\ref{tab:temps}. By eye there is no difference between the three datasets, and in all cases there is a clear dependence of both $\Omega_{DE}$ and $w$ on the depth of the Spot. A Hot Spot, which {\color{mydarkdarkred} in the confined parameterisation of this paper} is an under-density, tends to drive both $\Omega_{DE}$ and $w$ down, and a Cold Spot has the opposite behaviour.

The effect on $\Omega_{DE}$ may be surprising, since an under-density mimics $\Omega_{DE}$, such that one could expect that these mock supernovae ask for a larger value of $\Omega_{DE}$. However, the radius of 600 Mpc extends to a redshift of roughly $z\sim 0.14$. This means that only very nearby a higher value of $\Omega_{DE}$ is needed, which pushes $w$ down, while at the same time at higher redshifts the shape of the luminosity-distance-redshift relation is unchanged with respect to the $\Lambda$CDM curve. Since $\Omega_{DE}$ can be thought of roughly as the first derivative of the luminosity-distance-redshift curve, and $w$ as the second derivative, the two can slightly compensate each other, depending on the size of error bars. A more negative $w$ is hence compensated by a smaller value of $\Omega_{DE}$, in order to stay in agreement with the high redshift supernovae. For a Cold Spot (over-density), the converse holds.

In Table~\ref{tab:res} we show the numerical values of the marginalized one-dimensional posterior probabilities for the mock data based on the Union2 dataset, and we display the same results graphically in Figure~\ref{fig:wlambda}, emphasizing the relation between perceived $\Omega_{DE}$ and $w$ from the data on one hand, and $\Delta T / T$ that describes the object with which we created the mock data on the other hand. Considering this figure one can conclude that a Hot or Cold Spot with temperature deviation $T \sim \mathcal{O}(50$ $\mu$K$)$ affects $w$ at least at the ten percent level, baring in mind that the density perturbation in the Spot is over-estimated by the assumption of full matter domination at photon decoupling.

\begin{figure}
\begin{tabular}{ccc}
\includegraphics[width=0.3\textwidth]{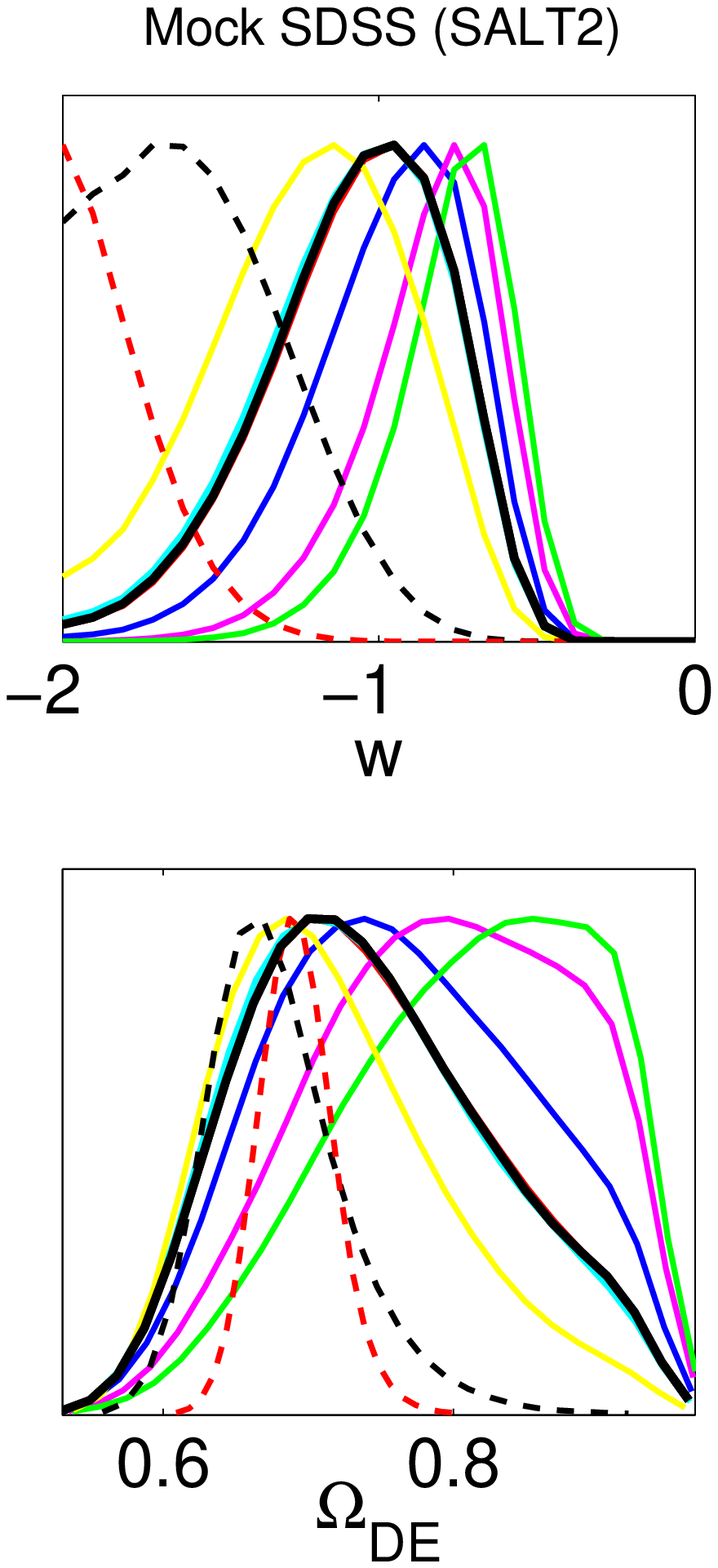}&
\includegraphics[width=0.3\textwidth]{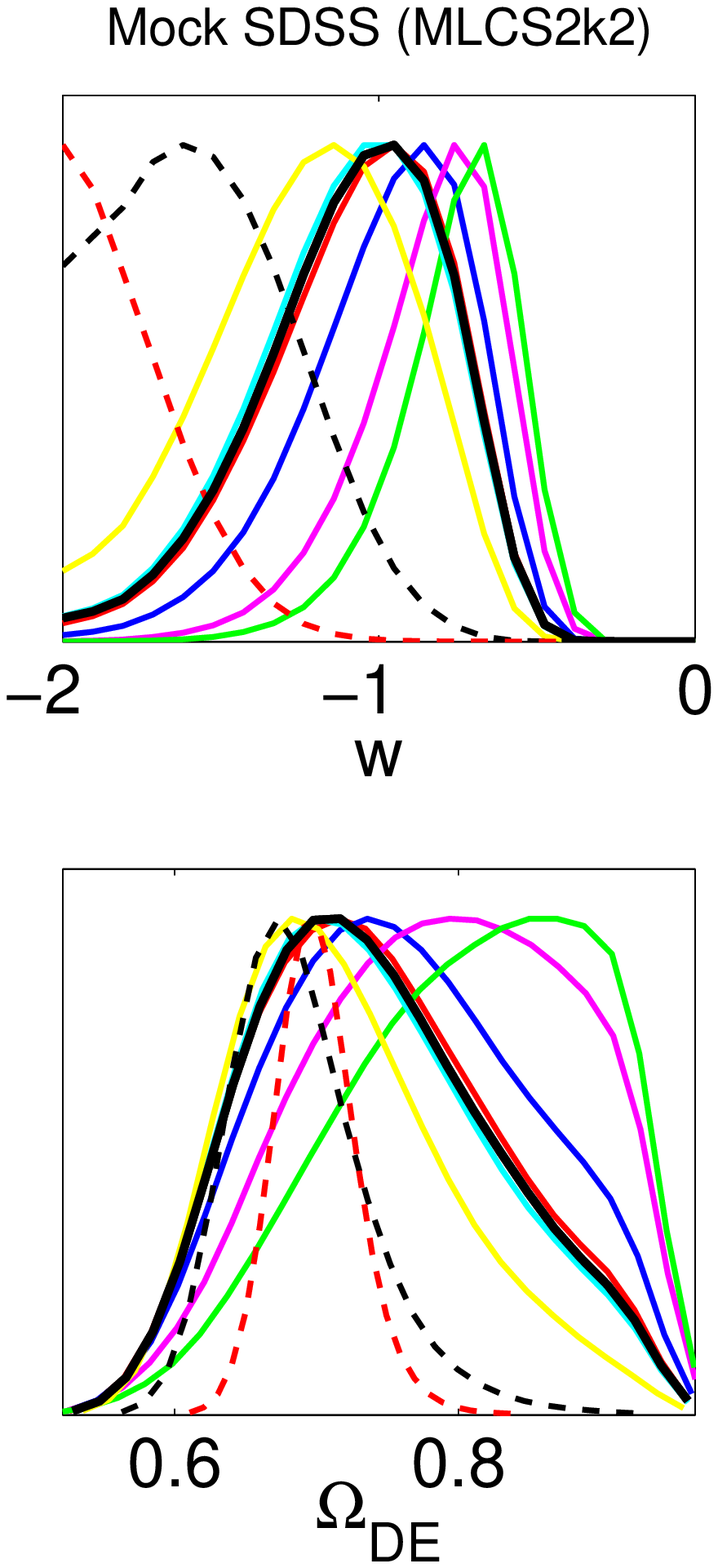}&
\includegraphics[width=0.3\textwidth]{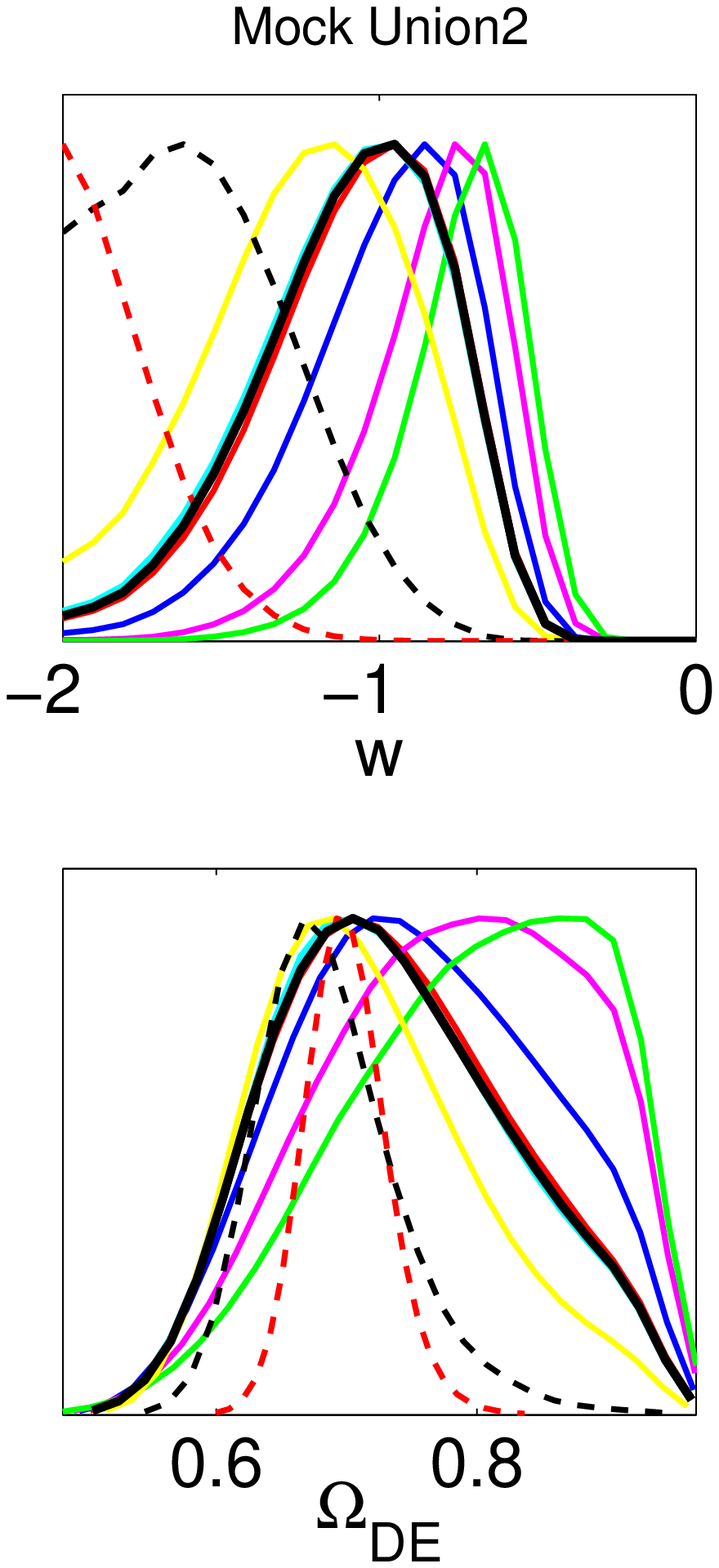}
\end{tabular}
\caption{Marginalized 1D posterior parameter likelihoods for $\Omega_DE$ and $w$ when fit to mock data based on the SDSS SN data with the SALT-II light curve fitter (left) and the MLCS2k2 light curve fitter (second from left), fit to mock data based on the Union2 sample with systematic errors (second form right) and without systematic errors (right). In each figure, the central thick black line corresponds to a pure $\Lambda$CDM cosmology mock data set. For both $w$ and $DE$, the likelihood that peaks at the left most value corresponds to the fit to a mock dataset based on an observer living in a Hot Spot of 177.69 $\mu$K (red dashed line) and the likelihood that peaks at the right most value corresponds to the fit to a mock dataset based on an observer living in a Cold Spot of  -152.63 $\mu$K (green solid line). All the other lines correspond to all the Spots listed in Table~\ref{tab:temps}, where hence going from top to bottom in that table correspond to the peaks in this figure going from right to left. }\label{fig:1dpostcurv} 
\end{figure}

\setlength{\mystretch}{0.5em}
\begin{table}
\begin{tabular*}{\textwidth}{@{\extracolsep{\fill}}rrr}
Temperature of Spot ($\mu$K) & $\Omega_{DE}$ & $w$\\ \hline
-127.966 & $ 0.80^{0.13}_{-0.20} $ & $ -0.74^{0.21}_{-0.35} $ \\[\mystretch]
-95.087 & $ 0.78^{0.15}_{-0.19} $ & $ -0.83^{0.25}_{-0.41} $ \\[\mystretch]
-40.633 & $ 0.76^{0.17}_{-0.18} $ & $ -0.98^{0.33}_{-0.50} $ \\[\mystretch]
-4.611 & $ 0.74^{0.18}_{-0.16} $ & $ -1.08^{0.39}_{-0.53} $ \\[\mystretch]
0.000 & $ 0.74^{0.18}_{-0.16} $ & $ -1.09^{0.39}_{-0.53} $ \\[\mystretch]
4.749 & $ 0.73^{0.19}_{-0.15} $ & $ -1.11^{0.41}_{-0.54} $ \\[\mystretch]
54.666 & $ 0.72^{0.18}_{-0.14} $ & $ -1.23^{0.46}_{-0.77} $ \\[\mystretch]
237.189 & $ 0.69^{0.13}_{-0.09} $ & $ -1.56^{0.49}_{-0.44} $ \\[\mystretch]
645.131 & $ 0.70^{0.07}_{-0.06} $ & $ -1.80^{0.32}_{-0.20} $ \\[\mystretch]
\end{tabular*}
\caption{Marginalized posterior likelihoods of $\Omega_{DE}$ and $w$ for an observer living in Hot and Cold Spots of different temperatures, all with a diameter of $5^\circ$. The Spot of $0$ $\mu$K corresponds to exact $\Lambda$CDM. The error bars are at 95\% confidence level.}\label{tab:res}
\end{table}

\section{Conclusion\label{sec:conc}}
We have for the first time made a link between the observation of large and `homogeneous' Cold Spots on the CMB and the apparent magnitude of the equation of state of Dark Energy, $w$, by placing the observer in a Spot identical to the one he or she observes on the CMB. 

We have modeled exact density perturbations on the surface of last scattering, using the LTB metric with a cosmological constant, such that we can describe the perturbation at the epoch of last scattering and today. Placing us as the observer in the center of such a perturbation, with co-moving radius of 600 Mpc and putting an identical perturbation on the surface of last scattering, producing a Spot of 5$^\circ$, we find that $w$ can be changed by up to ten percent for $\Delta T \sim \mathcal{O}(50$ $\mu$K$)$, depending on the central temperature of the Spot. This result is best illustrated in Figure~\ref{fig:wlambda}.

We used the exact solutions to the $\Lambda$LTB framework in the module {\sc ColLambda}~\cite{Valkenburg:2011tm} combined with the {\sc VoidDistances} module as published in~\cite{Biswas:2010xm}.

The main assumptions that deserve further investigation for improvement of this analysis are: (1) the crude approximation that radiation plays no role in the background equations at the epoch of decoupling, (2) the assumption that the last scattering surface goes through the center of the spherical object that causes the Spot, and (3) the simplification that the observer lives at $r=0$. We expect that improvements in points (1) and (2) will adjust the size of the effect we found here by a small margin, although point (2) may change the direction of the effect entirely, as explained in Section~\ref{sec:shorts}. Placing the observer off-center most likely will drive the observed values of $w$ and $\Omega_\Lambda$ closer to their fundamental values, but in that case the object in which the observer lives will still induce a spread in observed distances like in the Swiss-Cheese scenario~\cite{Vanderveld:2008vi,Valkenburg:2009iw}, hence causing larger error bars on $w$ and $\Omega_\Lambda$

\section*{Acknowledgements}
The author wishes to express his gratitude to Alexei Starobinsky for proposing the idea that led to this work. It is a pleasure to thank Alessio Notari, Valerio Marra, Martin Kunz, Kimmo Kainulainen, Juan Garcia-Bellido and Yvonne Wong for useful discussions.
\bibliographystyle{JHEP}
\bibliography{refs}

\providecommand{\href}[2]{#2}\begingroup\raggedright\begin{thebibliography}{10}

\bibitem{Riess:1998cb}
{\bf Supernova Search Team} Collaboration, A.~G. Riess {\em et.~al.}, {\it
  {Observational Evidence from Supernovae for an Accelerating Universe and a
  Cosmological Constant}},  {\em Astron. J.} {\bf 116} (1998) 1009--1038,
  [\href{http://xxx.lanl.gov/abs/astro-ph/9805201}{{\tt astro-ph/9805201}}].
%%CITATION = ASTRO-PH/9805201;%%.

\bibitem{Perlmutter:1998np}
{\bf Supernova Cosmology Project} Collaboration, S.~Perlmutter {\em et.~al.},
  {\it {Measurements of Omega and Lambda from 42 High-Redshift Supernovae}},
  {\em Astrophys. J.} {\bf 517} (1999) 565--586,
  [\href{http://xxx.lanl.gov/abs/astro-ph/9812133}{{\tt astro-ph/9812133}}].
%%CITATION = ASTRO-PH/9812133;%%.

\bibitem{Wetterich:1987fm}
C.~Wetterich, {\it {Cosmology and the Fate of Dilatation Symmetry}},  {\em
  Nucl. Phys.} {\bf B302} (1988) 668.
%%CITATION = NUPHA,B302,668;%%.

\bibitem{Sahni:1999gb}
V.~Sahni and A.~A. Starobinsky, {\it {The Case for a Positive Cosmological
  Lambda-term}},  {\em Int. J. Mod. Phys.} {\bf D9} (2000) 373--444,
  [\href{http://xxx.lanl.gov/abs/astro-ph/9904398}{{\tt astro-ph/9904398}}].
%%CITATION = ASTRO-PH/9904398;%%.

\bibitem{1995ApJ...453...17M:1995}
J.~W. {Moffat} and D.~C. {Tatarski}, {\it {Cosmological Observations in a Local
  Void}},  {\em Astrophys. J.} {\bf 453} (Nov., 1995) 17--+,
[\href{http://xxx.lanl.gov/abs/astro-ph/}{{\tt astro-ph/}}].

\bibitem{Tomita:1999qn}
K.~Tomita, {\it {Distances and lensing in cosmological void models}},  {\em
  Astrophys. J.} {\bf 529} (2000) 38,
  [\href{http://xxx.lanl.gov/abs/astro-ph/9906027}{{\tt astro-ph/9906027}}].
%%CITATION = ASTRO-PH/9906027;%%.

\bibitem{Mansouri:2005rf}
R.~Mansouri, {\it {Structured FRW universe leads to acceleration: A non-
  perturbative approach}},
  \href{http://xxx.lanl.gov/abs/astro-ph/0512605}{{\tt astro-ph/0512605}}.
%%CITATION = ASTRO-PH/0512605;%%.

\bibitem{Alnes:2005rw}
H.~Alnes, M.~Amarzguioui, and O.~Gron, {\it {An inhomogeneous alternative to
  dark energy?}},  {\em Phys. Rev.} {\bf D73} (2006) 083519,
  [\href{http://xxx.lanl.gov/abs/astro-ph/0512006}{{\tt astro-ph/0512006}}].
%%CITATION = ASTRO-PH/0512006;%%.

\bibitem{Enqvist:2006cg}
K.~Enqvist and T.~Mattsson, {\it {The effect of inhomogeneous expansion on the
  supernova observations}},  {\em JCAP} {\bf 0702} (2007) 019,
  [\href{http://xxx.lanl.gov/abs/astro-ph/0609120}{{\tt astro-ph/0609120}}].
%%CITATION = ASTRO-PH/0609120;%%.

\bibitem{Enqvist:2007vb}
K.~Enqvist, {\it {Lemaitre-Tolman-Bondi model and accelerating expansion}},
  {\em Gen. Rel. Grav.} {\bf 40} (2008) 451--466,
  [\href{http://xxx.lanl.gov/abs/0709.2044}{{\tt arXiv:0709.2044}}].
%%CITATION = 0709.2044;%%.

\bibitem{GarciaBellido:2008nz}
J.~Garcia-Bellido and T.~Haugboelle, {\it {Confronting Lemaitre-Tolman-Bondi
  models with Observational Cosmology}},  {\em JCAP} {\bf 0804} (2008) 003,
  [\href{http://xxx.lanl.gov/abs/0802.1523}{{\tt arXiv:0802.1523}}].
%%CITATION = 0802.1523;%%.

\bibitem{Clifton:2009kx}
T.~Clifton, P.~G. Ferreira, and J.~Zuntz, {\it {What the small angle CMB really
  tells us about the curvature of the Universe}},  {\em JCAP} {\bf 0907} (2009)
  029, [\href{http://xxx.lanl.gov/abs/0902.1313}{{\tt arXiv:0902.1313}}].
%%CITATION = 0902.1313;%%.

\bibitem{February:2009pv}
S.~February, J.~Larena, M.~Smith, and C.~Clarkson, {\it {Rendering Dark Energy
  Void}},  {\em Mon. Not. Roy. Astron. Soc.} {\bf 405} (2010) 2231,
  [\href{http://xxx.lanl.gov/abs/0909.1479}{{\tt arXiv:0909.1479}}].
%%CITATION = 0909.1479;%%.

\bibitem{Quartin:2009xr}
M.~Quartin and L.~Amendola, {\it {Distinguishing Between Void Models and Dark
  Energy with Cosmic Parallax and Redshift Drift}},  {\em Phys. Rev.} {\bf D81}
  (2010) 043522, [\href{http://xxx.lanl.gov/abs/0909.4954}{{\tt
  arXiv:0909.4954}}].
%%CITATION = 0909.4954;%%.

\bibitem{Marra:2010pg}
V.~Marra and M.~Paakkonen, {\it {Observational constraints on the LLTB model}},
   {\em JCAP} {\bf 1012} (2010) 021,
  [\href{http://xxx.lanl.gov/abs/1009.4193}{{\tt arXiv:1009.4193}}].
%%CITATION = 1009.4193;%%.

\bibitem{Moss:2010jx}
A.~Moss, J.~P. Zibin, and D.~Scott, {\it {Precision cosmology defeats void
  models for acceleration}},  {\em Phys. Rev.} {\bf D83} (2011) 103515,
  [\href{http://xxx.lanl.gov/abs/1007.3725}{{\tt arXiv:1007.3725}}].
%%CITATION = 1007.3725;%%.

\bibitem{Biswas:2010xm}
T.~Biswas, A.~Notari, and W.~Valkenburg, {\it {Testing the Void against
  Cosmological data: fitting CMB, BAO, SN and H0}},  {\em JCAP} {\bf 1011}
  (2010) 030, [\href{http://xxx.lanl.gov/abs/1007.3065}{{\tt
  arXiv:1007.3065}}].
%%CITATION = 1007.3065;%%.

\bibitem{Nadathur:2010zm}
S.~Nadathur and S.~Sarkar, {\it {Reconciling the local void with the CMB}},
  {\em Phys. Rev.} {\bf D83} (2011) 063506,
  [\href{http://xxx.lanl.gov/abs/1012.3460}{{\tt arXiv:1012.3460}}].
%%CITATION = 1012.3460;%%.

\bibitem{Marra:2011ct}
V.~Marra and A.~Notari, {\it {Observational constraints on inhomogeneous
  cosmological models without dark energy}},  {\em Class. Quant. Grav.} {\bf
  28} (2011) 164004, [\href{http://xxx.lanl.gov/abs/1102.1015}{{\tt
  arXiv:1102.1015}}].
%%CITATION = 1102.1015;%%.

\bibitem{Bolejko:2011jc}
K.~Bolejko, M.-N. Celerier, and A.~Krasinski, {\it {Inhomogeneous cosmological
  models: exact solutions and their applications}},  {\em Class. Quant. Grav.}
  {\bf 28} (2011) 164002, [\href{http://xxx.lanl.gov/abs/1102.1449}{{\tt
  arXiv:1102.1449}}].
%%CITATION = 1102.1449;%%.

\bibitem{Sussman:2011na}
R.~A. Sussman, {\it {Back-reaction and effective acceleration in generic LTB
  dust models}},  {\em Class. Quant. Grav.} {\bf 28} (2011) 235002,
  [\href{http://xxx.lanl.gov/abs/1102.2663}{{\tt arXiv:1102.2663}}].
%%CITATION = 1102.2663;%%.

\bibitem{Valkenburg:2009iw}
W.~Valkenburg, {\it {Swiss Cheese and a Cheesy CMB}},  {\em JCAP} {\bf 0906}
  (2009) 010, [\href{http://xxx.lanl.gov/abs/0902.4698}{{\tt
  arXiv:0902.4698}}].
%%CITATION = 0902.4698;%%.

\bibitem{Cruz:2006sv}
M.~Cruz, M.~Tucci, E.~Martinez-Gonzalez, and P.~Vielva, {\it {The non-Gaussian
  Cold Spot in WMAP: significance, morphology and foreground contribution}},
  {\em Mon. Not. Roy. Astron. Soc.} {\bf 369} (2006) 57--67,
  [\href{http://xxx.lanl.gov/abs/astro-ph/0601427}{{\tt astro-ph/0601427}}].
%%CITATION = ASTRO-PH/0601427;%%.

\bibitem{Zhang:2009qg}
R.~Zhang and D.~Huterer, {\it {Disks in the sky: A reassessment of the WMAP
  'cold spot'}},  {\em Astropart. Phys.} {\bf 33} (2010) 69--74,
  [\href{http://xxx.lanl.gov/abs/0908.3988}{{\tt arXiv:0908.3988}}].
%%CITATION = 0908.3988;%%.

\bibitem{Bennett:2010jb}
C.~L. Bennett {\em et.~al.}, {\it {Seven-Year Wilkinson Microwave Anisotropy
  Probe (WMAP) Observations: Are There Cosmic Microwave Background
  Anomalies?}},  {\em Astrophys. J. Suppl.} {\bf 192} (2011) 17,
  [\href{http://xxx.lanl.gov/abs/1001.4758}{{\tt arXiv:1001.4758}}].
%%CITATION = 1001.4758;%%.

\bibitem{Tomita:2005nu}
K.~Tomita, {\it {Second-order gravitational effects of local inhomogeneities on
  CMB anisotropies and non-Gaussian signatures}},  {\em Phys. Rev.} {\bf D72}
  (2005) 103506, [\href{http://xxx.lanl.gov/abs/astro-ph/0509518}{{\tt
  astro-ph/0509518}}].
%%CITATION = ASTRO-PH/0509518;%%.

\bibitem{Inoue:2006rd}
K.~T. Inoue and J.~Silk, {\it {Local Voids as the Origin of Large-angle Cosmic
  Microwave Background Anomalies}},  {\em Astrophys. J.} {\bf 648} (2006)
  23--30, [\href{http://xxx.lanl.gov/abs/astro-ph/0602478}{{\tt
  astro-ph/0602478}}].
%%CITATION = ASTRO-PH/0602478;%%.

\bibitem{Inoue:2006fn}
K.~T. Inoue and J.~Silk, {\it {Local Voids as the Origin of Large-angle Cosmic
  Microwave Background Anomalies: The Effect of a Cosmological Constant}},
  {\em Astrophys. J.} {\bf 664} (2007) 650--659,
  [\href{http://xxx.lanl.gov/abs/astro-ph/0612347}{{\tt astro-ph/0612347}}].
%%CITATION = ASTRO-PH/0612347;%%.

\bibitem{Masina:2008zv}
I.~Masina and A.~Notari, {\it {The Cold Spot as a Large Void: Rees-Sciama
  effect on CMB Power Spectrum and Bispectrum}},  {\em JCAP} {\bf 0902} (2009)
  019, [\href{http://xxx.lanl.gov/abs/0808.1811}{{\tt arXiv:0808.1811}}].
%%CITATION = 0808.1811;%%.

\bibitem{Ayaita:2009xm}
Y.~Ayaita, M.~Weber, and C.~Wetterich, {\it {Too few spots in the Cosmic
  Microwave Background}},  {\em Phys. Rev.} {\bf D81} (2010) 023507,
  [\href{http://xxx.lanl.gov/abs/0905.3324}{{\tt arXiv:0905.3324}}].
%%CITATION = 0905.3324;%%.

\bibitem{Komatsu:2010fb}
{\bf WMAP} Collaboration, E.~Komatsu {\em et.~al.}, {\it {Seven-Year Wilkinson
  Microwave Anisotropy Probe (WMAP) Observations: Cosmological
  Interpretation}},  {\em Astrophys. J. Suppl.} {\bf 192} (2011) 18,
  [\href{http://xxx.lanl.gov/abs/1001.4538}{{\tt arXiv:1001.4538}}].
%%CITATION = 1001.4538;%%.

\bibitem{Amendola:2010ub}
L.~Amendola, K.~Kainulainen, V.~Marra, and M.~Quartin, {\it {Large-scale
  inhomogeneities may improve the cosmic concordance of supernovae}},  {\em
  Phys. Rev. Lett.} {\bf 105} (2010) 121302,
  [\href{http://xxx.lanl.gov/abs/1002.1232}{{\tt arXiv:1002.1232}}].
%%CITATION = 1002.1232;%%.

\bibitem{Sinclair:2010sb}
B.~Sinclair, T.~M. Davis, and T.~Haugbolle, {\it {Residual Hubble-bubble
  effects on supernova cosmology}},  {\em Astrophys. J.} {\bf 718} (2010)
  1445--1455, [\href{http://xxx.lanl.gov/abs/1006.0911}{{\tt
  arXiv:1006.0911}}].
%%CITATION = 1006.0911;%%.

\bibitem{Romano:2010nc}
A.~E. Romano, M.~Sasaki, and A.~A. Starobinsky, {\it {Effects of
  inhomogeneities on apparent cosmological observables: 'fake'' evolving dark
  energy}},  \href{http://xxx.lanl.gov/abs/1006.4735}{{\tt arXiv:1006.4735}}.
%%CITATION = 1006.4735;%%.

\bibitem{Romano:2011mx}
A.~E. Romano and P.~Chen, {\it {Corrections to the apparent value of the
  cosmological constant due to local inhomogeneities}},  {\em JCAP} {\bf 1110}
  (2011) 016, [\href{http://xxx.lanl.gov/abs/1104.0730}{{\tt
  arXiv:1104.0730}}].
%%CITATION = 1104.0730;%%.

\bibitem{deLavallaz:2011tj}
A.~de~Lavallaz and M.~Fairbairn, {\it {Effects of voids on the reconstruction
  of the equation of state of Dark Energy}},  {\em Phys. Rev.} {\bf D84} (2011)
  083005, [\href{http://xxx.lanl.gov/abs/1106.1611}{{\tt arXiv:1106.1611}}].
%%CITATION = 1106.1611;%%.

\bibitem{Valkenburg:2011tm}
W.~Valkenburg, {\it {Exact spherical collapse of dust in an expanding spacetime
  with a cosmological constant and solutions to the Friedmann equation}},
  \href{http://xxx.lanl.gov/abs/1104.1082}{{\tt arXiv:1104.1082}}.
%%CITATION = 1104.1082;%%.

\bibitem{1995NuAlg..10...13C}
B.~C. {Carlson}, {\it {Numerical computation of real or complex elliptic
  integrals}},  {\em Numerical Algorithms} {\bf 10} (Mar., 1995) 13--26,
[\href{http://xxx.lanl.gov/abs/math/9409}{{\tt math/9409}}].

\bibitem{Biswas:2007gi}
T.~Biswas and A.~Notari, {\it {Swiss-Cheese Inhomogeneous Cosmology \& the Dark
  Energy Problem}},  {\em JCAP} {\bf 0806} (2008) 021,
  [\href{http://xxx.lanl.gov/abs/astro-ph/0702555}{{\tt astro-ph/0702555}}].
%%CITATION = ASTRO-PH/0702555;%%.

\bibitem{VanAcoleyen:2008cy}
K.~Van~Acoleyen, {\it {LTB solutions in Newtonian gauge: from strong to weak
  fields}},  {\em JCAP} {\bf 0810} (2008) 028,
  [\href{http://xxx.lanl.gov/abs/0808.3554}{{\tt arXiv:0808.3554}}].
%%CITATION = 0808.3554;%%.

\bibitem{Sachs:1967er}
R.~K. Sachs and A.~M. Wolfe, {\it {Perturbations of a cosmological model and
  angular variations of the microwave background}},  {\em Astrophys. J.} {\bf
  147} (1967) 73--90.
%%CITATION = ASJOA,147,73;%%.

\bibitem{Durrer:2001gq}
R.~Durrer, {\it {The theory of CMB anisotropies}},  {\em J. Phys. Stud.} {\bf
  5} (2001) 177--215, [\href{http://xxx.lanl.gov/abs/astro-ph/0109522}{{\tt
  astro-ph/0109522}}].
%%CITATION = ASTRO-PH/0109522;%%.

\bibitem{Bennett:1996ce}
C.~L. Bennett {\em et.~al.}, {\it {4-Year COBE DMR Cosmic Microwave Background
  Observations: Maps and Basic Results}},  {\em Astrophys. J.} {\bf 464} (1996)
  L1--L4, [\href{http://xxx.lanl.gov/abs/astro-ph/9601067}{{\tt
  astro-ph/9601067}}].
%%CITATION = ASTRO-PH/9601067;%%.

\bibitem{Granett:2008ju}
B.~R. Granett, M.~C. Neyrinck, and I.~Szapudi, {\it {An Imprint of
  Super-Structures on the Microwave Background due to the Integrated
  Sachs-Wolfe Effect}},  \href{http://xxx.lanl.gov/abs/0805.3695}{{\tt
  arXiv:0805.3695}}.
%%CITATION = 0805.3695;%%.

\bibitem{Inoue:2010rp}
K.~T. Inoue, N.~Sakai, and K.~Tomita, {\it {Evidence of Quasi-linear
  Super-Structures in the Cosmic Microwave Background and Galaxy
  Distribution}},  {\em Astrophys. J.} {\bf 724} (2010) 12--25,
  [\href{http://xxx.lanl.gov/abs/1005.4250}{{\tt arXiv:1005.4250}}].
%%CITATION = 1005.4250;%%.

\bibitem{Park:2010xw}
J.~Park, C.-G. Park, and J.-c. Hwang, {\it {Analysis of recent type Ia
  supernova data based on evolving dark energy models}},  {\em Phys. Rev.} {\bf
  D84} (2011) 023506, [\href{http://xxx.lanl.gov/abs/1011.1723}{{\tt
  arXiv:1011.1723}}].
%%CITATION = 1011.1723;%%.

\bibitem{Kessler:2009ys}
R.~Kessler {\em et.~al.}, {\it {First-year Sloan Digital Sky Survey-II
  (SDSS-II) Supernova Results: Hubble Diagram and Cosmological Parameters}},
  {\em Astrophys. J. Suppl.} {\bf 185} (2009) 32--84,
  [\href{http://xxx.lanl.gov/abs/0908.4274}{{\tt arXiv:0908.4274}}].
%%CITATION = 0908.4274;%%.

\bibitem{Amanullah:2010vv}
R.~Amanullah {\em et.~al.}, {\it {Spectra and Light Curves of Six Type Ia
  Supernovae at 0.511 < z < 1.12 and the Union2 Compilation}},  {\em Astrophys.
  J.} {\bf 716} (2010) 712--738, [\href{http://xxx.lanl.gov/abs/1004.1711}{{\tt
  arXiv:1004.1711}}].
%%CITATION = 1004.1711;%%.

\bibitem{Lewis:2002ah}
A.~Lewis and S.~Bridle, {\it {Cosmological parameters from CMB and other data:
  a Monte- Carlo approach}},  {\em Phys. Rev.} {\bf D66} (2002) 103511,
  [\href{http://xxx.lanl.gov/abs/astro-ph/0205436}{{\tt astro-ph/0205436}}].
%%CITATION = ASTRO-PH/0205436;%%.

\bibitem{Lewis:1999bs}
A.~Lewis, A.~Challinor, and A.~Lasenby, {\it {Efficient Computation of CMB
  anisotropies in closed FRW models}},  {\em Astrophys. J.} {\bf 538} (2000)
  473--476, [\href{http://xxx.lanl.gov/abs/astro-ph/9911177}{{\tt
  astro-ph/9911177}}].
%%CITATION = ASTRO-PH/9911177;%%.

\bibitem{Vanderveld:2008vi}
R.~A. Vanderveld, E.~E. Flanagan, and I.~Wasserman, {\it {Luminosity distance
  in 'Swiss cheese' cosmology with randomized voids: I. Single void size}},
  {\em Phys. Rev.} {\bf D78} (2008) 083511,
  [\href{http://xxx.lanl.gov/abs/0808.1080}{{\tt arXiv:0808.1080}}].
%%CITATION = 0808.1080;%%.

\end{thebibliography}\endgroup

\end{document}